\documentclass[aps,prd,twocolumn,superscriptaddress, nofootinbib]{revtex4-1}
\usepackage{epsf,epsfig,graphicx}
\usepackage{amsmath}
\usepackage{amssymb}
\usepackage{tikz}
\usepackage{subfigure}
\usepackage{multirow}
\usepackage{tabularx}
\makeatletter
\DeclareFontFamily{OMX}{MnSymbolE}{}
\DeclareSymbolFont{MnLargeSymbols}{OMX}{MnSymbolE}{m}{n}
\SetSymbolFont{MnLargeSymbols}{bold}{OMX}{MnSymbolE}{b}{n}
\DeclareFontShape{OMX}{MnSymbolE}{m}{n}{
    <-6>  MnSymbolE5
   <6-7>  MnSymbolE6
   <7-8>  MnSymbolE7
   <8-9>  MnSymbolE8
   <9-10> MnSymbolE9
  <10-12> MnSymbolE10
  <12->   MnSymbolE12
}{}
\DeclareFontShape{OMX}{MnSymbolE}{b}{n}{
    <-6>  MnSymbolE-Bold5
   <6-7>  MnSymbolE-Bold6
   <7-8>  MnSymbolE-Bold7
   <8-9>  MnSymbolE-Bold8
   <9-10> MnSymbolE-Bold9
  <10-12> MnSymbolE-Bold10
  <12->   MnSymbolE-Bold12
}{}

\let\llangle\@undefined
\let\rrangle\@undefined
\DeclareMathDelimiter{\llangle}{\mathopen}%
                     {MnLargeSymbols}{'164}{MnLargeSymbols}{'164}
\DeclareMathDelimiter{\rrangle}{\mathclose}%
                     {MnLargeSymbols}{'171}{MnLargeSymbols}{'171}
\makeatother

\RequirePackage[colorlinks,citecolor=red,urlcolor=red,linkcolor=red]{hyperref}

\def\edth{\;\raise1.0pt\hbox{$'$}\hskip-6pt\partial\;}
\def\baredth{\;\overline{\raise1.0pt\hbox{$'$}\hskip-6pt
\partial}\;}
\def\gsim{~\rlap{$>$}{\lower 1.0ex\hbox{$\sim$}}}

\newcommand{\be}{\begin{equation}}
\newcommand{\ee}{\end{equation}}
\newcommand{\bw}{\begin{widetext}}
\newcommand{\ew}{\end{widetext}}

\newcommand{\intzeroinf}{\int_{0}^{\infty}}

\usepackage{xcolor}

\usepackage[normalem]{ulem}

\AtBeginDocument{}

\begin{document}

\title{Accurate pulsar timing array residual variances and correlation \\
of the stochastic gravitational wave background}

\author{Reginald Christian Bernardo}
\email{reginald.christian.bernardo@aei.mpg.de}
\affiliation{Max Planck Institute for Gravitational Physics (Albert Einstein Institute), Hannover D-30167, Germany}
\affiliation{Asia Pacific Center for Theoretical Physics, Pohang 37673, Korea}

\author{Kin-Wang Ng}
\email{nkw@phys.sinica.edu.tw}
\affiliation{Institute of Physics, Academia Sinica, Taipei 11529, Taiwan}
\affiliation{Institute of Astronomy and Astrophysics, Academia Sinica, Taipei 11529, Taiwan}


\begin{abstract}
Pulsar timing arrays have reported a compelling evidence of a nanohertz stochastic gravitational wave background. However, the origin of the signal remains undetermined, largely because its spectrum is bluer for an astrophysical source and can be explained by cosmological models. In this letter, we {revisit the frequency- and Fourier-domain analysis of the signal by} {deriving} theoretically accurate expressions for the Fourier bin variances and correlation of pulsar timing residuals, and demonstrate their outstanding agreement with {point source} astrophysical simulations. In contrast, we show that a common power law {(or a diagonal covariance approximation)} traditionally used to interpret a stochastic gravitational wave background {signal} is generally faced with systematic errors, one of which is the illusion of a bluer signal. This hints at a conservative solution, supportive of an astrophysical source, to the observed correlated common spectrum process in pulsar timing arrays.
\end{abstract}

\maketitle

Pulsar timing array (PTA) collaborations have provided evidence of a correlated common spectrum process across Galactic millisecond pulsars, consistent with a stochastic gravitational wave background (SGWB) in the nanohertz regime \cite{NANOGrav:2023gor, Reardon:2023gzh, EPTA:2023fyk, Xu:2023wog}. This signal is characterized by a quadrupolar correlation matching the long-sought Hellings \& Downs (HD) curve \cite{Hellings:1983fr}. However, the constrained spectrum appears bluer than expected compared with the conservative supermassive black hole binaries (SMBHB) model, sparking a tantalizing debate about the origin of the signal \cite{Phinney:2001di, Sesana:2004sp, EPTA:2023xxk, NANOGrav:2023hvm, Wu:2023hsa, Ellis:2023oxs, Bian:2023dnv, Bi:2023tib, Wang:2023ost, Wang:2023sij, Verbiest:2024nid}.

This situation has propelled the field in several directions. One possibility is that current PTA data may be insufficient to definitively distinguish between various potential astrophysical and cosmological sources, suggesting that stronger constraints might emerge only with future data \cite{Lazio:2013mea, Weltman:2018zrl}. Alternatively, we can explore new theoretical aspects, such as cosmic variance \cite{Allen:2022dzg, Bernardo:2022xzl}, or observational techniques like analyzing the cumulants of the spectrum \cite{Lamb:2024gbh} and correlation \cite{Bernardo:2024uiq}, to fully leverage the capability of PTAs in constraining physical models.

Now is also an opportune time to revisit the foundational elements of both theory and PTA data analysis, addressing any systematic errors that could obscure or exaggerate the detection of a SGWB signal \cite{Allen:2024uqs, DiMarco:2024irz, Goncharov:2024htb, Crisostomi:2025vue, Bernardo:2025dat}. This letter is geared toward this direction---reexamining the interpretation of a SGWB signal {in the frequency- and Fourier-domain} \cite{Valtolina:2024kdb} of the PTA timing residuals, leading to new accurate expressions for the residuals' bin variances and correlation (\ref{eq:xy_correlation}-\ref{eq:bb_filter}). The new expressions turns out to be remarkably consistent with astrophysical simulations \cite{Hobbs:2009yn}. Most importantly, this reexamination has brought to light facets of the standard analysis {built on a diagonal covariance approximation} that reflect as systematic errors on an inferred SGWB spectrum, such as biasing a spectrum to the blue side (Table \ref{tab:constraints}). 

This letter proceeds with a derivation of theoretically accurate PTA residual variances and correlation due to a SGWB, followed by a rigorous comparison with simulations. We use units $G=c=1$, subscripts $a,b$ to denote pulsars, and $j,k$ for frequency bins.

In practice, the pulsar timing residual is expanded as a Fourier series \cite{Taylor:2021yjx},
\be
\label{eq:timing_residual_fourier_series}
r(t,\hat{e})= {\alpha_0(\hat{e})\over2}+\sum_{k=1}^{\infty} \alpha_k(\hat{e}) \sin(\omega_k t) + \sum_{k=1}^{\infty} \beta_k(\hat{e}) \cos(\omega_k t)\,,
\ee
where $\omega_k = 2 \pi f_k$, $f_k=k/T$, $k$ is a positive integer, $T$ is the total observation time, and ${\hat e}$ is a pulsar direction unit vector\footnote{
The Fourier components of the timing residual are given by $\alpha_0(\hat{e})= (2/T)\int_0^T dt\, r(t,\hat{e})$, $\alpha_i(\hat{e})= (2/T)\int_0^T dt\, r(t,\hat{e}) \sin(\omega_i t)$, and $\beta_i(\hat{e})= (2/T)\int_0^T dt\, r(t,\hat{e}) \cos(\omega_i t)$.
}. For brevity, we write down $\langle X_{aj}Y_{bk} \rangle\equiv\langle X_{j}(\hat{e}_a) Y_{k}(\hat{e}_b) \rangle$ where $X_j(\hat{e}_a)$ and $Y_k(\hat{e}_b)$ represent either one of the timing residual's Fourier components $\alpha_j(\hat{e}_a)$ and $\beta_k(\hat{e}_b)$. Then, for the GWB part, it is taken that \cite{NANOGrav:2023icp, NANOGrav:2024xhc}
\begin{align}
\label{eq:PTA_ab_correlation}
\langle \alpha_{aj} \alpha_{bk} \rangle
&\propto I(f_k)\, \gamma^{\rm HD}(\hat{e}_a\cdot \hat{e}_b)\, \delta_{jk}\,,\nonumber \\
\langle \beta_{aj} \beta_{bk} \rangle &\propto I(f_k)\, \gamma^{\rm HD}(\hat{e}_a\cdot \hat{e}_b)\, \delta_{jk}\,,\nonumber \\
\langle \alpha_{aj} \beta_{bk} \rangle&= 0\,,
\end{align}
where $\gamma^{\rm HD}(\hat{e}_a\cdot \hat{e}_b)$ is the HD curve, $I(f)$ is the SGWB power spectrum, and $\langle \cdots \rangle$ denote ensemble averaging. {This follows by using a diagonal approximation to the covariance function \cite{Crisostomi:2025vue}.} If $T\sim1$ yr, then the frequency is $f_1 \sim 31.8$ nHz; if $T\sim15$ yr, then $f_1\sim 2.1$ nHz. Following this logic, PTAs are able to measure the power spectrum at lower frequencies, $\sim I(1/T)$, through longer observing periods and increasing sky coverage. The Kronecker delta, $\delta_{kj}$, can be read as an extremely localized filter at frequencies $\sim f_k$, giving PTAs source information at infinitesimally narrow bins \cite{Allen:2024uqs}.

In the following, we shall derive accurate PTA residual variances and correlation, as opposed to the highly idealized \eqref{eq:PTA_ab_correlation}.

Consider an isotropic and Gaussian SGWB. In symbols, we write down
\begin{equation}
\label{eq:I_def}
\begin{split}
    \langle h_A (f, \hat{k}) h_{A'}^*(f', \hat{k}') \rangle = \delta_{AA'} \delta(f - f') \delta(\hat{k} - \hat{k}') I(f) \,,
\end{split}
\end{equation}
where $h_A(f, \hat{k})$ are GW amplitudes. For simplicity, we consider equidistant pulsars relative to Earth, $D_a \sim D_b \sim D \sim 1$ kpc. Then, the timing residual cross correlation can be written as \cite{Liu:2022skj, Bernardo:2022rif, Depta:2024ykq}
\be
\label{eq:rarb_correlation}
\begin{split}
& \langle r(t_1,\hat{e}_a)r(t_2,\hat{e}_b) \rangle=\\
& \ \ \ \ \ \ \ \ \ \ \ \intzeroinf\frac{2\,df}{(2\pi f)^2}  I(f) \gamma(fD,\hat{e}_a \cdot \hat{e}_b) C(f,t_1,t_2) \,,
\end{split}
\ee
where $\gamma(fD,\hat{e}_a \cdot \hat{e}_b)$ is the spatial correlation, 
and
\be
\label{eq:tt'_correlation}
C(f,t_1,t_2)= 1-\cos(2\pi f t_1) - \cos(2\pi f t_2) + \cos[2\pi f (t_2-t_1)]\,
\ee
is a temporal correlation owed to wave propagation in time. The spatial correlation can be written as
\begin{equation}
\label{eq:correlation_harmonic_series}
    \gamma(fD,\hat{e}_a \cdot \hat{e}_b) = \sum_{l \geq 2} \dfrac{2l+1}{4\pi} C_l(fD) P_l( \hat{e}_a \cdot \hat{e}_b ) \,,
\end{equation}
where the angular power spectrum, $C_l(y)$'s, is given by
\begin{equation}
\begin{split}
    C_l(y) = 2\pi^{3/2} \dfrac{(l+2)!}{(l-2)!} \bigg| \int_0^{2\pi y} dx \ e^{i x} \dfrac{j_l(x)}{x^2} \bigg|^2 \,.
\end{split}
\end{equation}
{T}he finite upper limit to the integral corresponds to the pulsar term. When dropped, $f D\gg1$, as in Galactic millisecond pulsars and nanohertz frequencies, the $C_l(y)$'s reduce to the HD correlation power spectrum, $C_l(\infty)\sim C_l^{\rm HD} = 8\pi^{3/2}/((l-1)l(l+1)(l+2))$; in real space, the sum \eqref{eq:correlation_harmonic_series} can be analytically closed to give rise to the HD curve, $\gamma^{\rm HD}(\hat{e}_a\cdot \hat{e}_b) = \gamma(\infty, \hat{e}_a\cdot \hat{e}_b)$. {The terms $1$, $\cos(2\pi ft_1)$, and $\cos(2f\pi t_2)$ in \eqref{eq:tt'_correlation} are a result of arbitrarily fixing the initial phase of the pulsar timing residuals; such that $r(t=0, \hat{e})=0$. In this light, (\ref{eq:rarb_correlation}-\ref{eq:tt'_correlation}) could be recognized as a statement of the Wiener-Khinchin theorem for the SGWB signal.}

\begin{figure}[t]
    \centering
    {\includegraphics[width=0.45\textwidth]{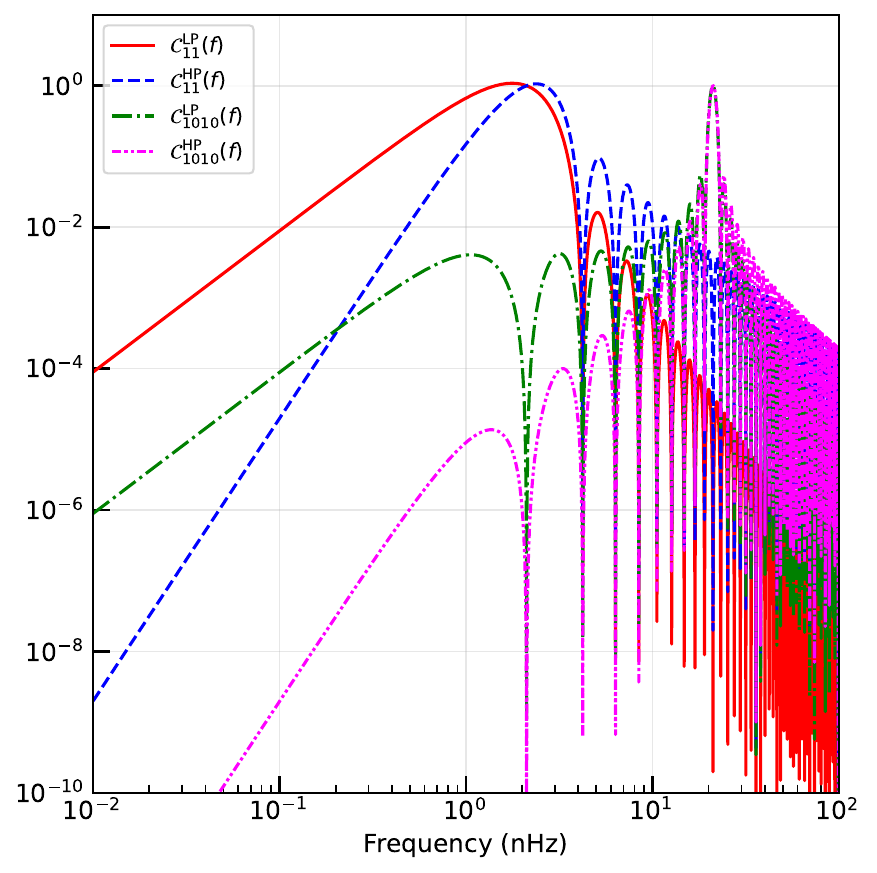}}
    \caption{Low-pass and high-pass {transfer functions} \eqref{eq:aa_filter} and \eqref{eq:bb_filter} in the 1st and 10th frequency bins with $T=15$ yr; $f_1 \sim 2$ nHz, $f_{10}=10 f_1\sim 20$ nHz.}
    \label{fig:filters}
\end{figure}

In fact, using \eqref{eq:rarb_correlation}, it can be shown that {\it all} the correlation components across frequency bins can be compactly expressed as
\begin{equation}
\label{eq:xy_correlation}
    \langle X_{aj} Y_{bk} \rangle = {\cal F}[ I(f), \gamma(fD,\hat{e}_a \cdot \hat{e}_b), \mathcal{C}_{jk}^{(XY)}(f) ] \,,
\end{equation}
where the functional ${\cal F}$ is given by
\begin{equation}
\label{eq:F_functional}
    {\cal F}[I(f), \gamma(fD, x), {\cal C}(f)] = \int_0^\infty \frac{2\,df}{(2\pi f)^2} I(f) \gamma(fD,x) {\cal C}(f) \,,
\end{equation}
and ${\cal C}_{jk}^{(XY)}(f)$ are {transfer functions} specific to every component of the correlation\footnote{
The timing residual correlation and its Fourier components can be read as a convolution between the power spectrum, the spatial correlation, and a {transfer function}.
}. For us, the relevant ones can be teased out analytically,
\begin{equation}
\label{eq:aa_filter}
\begin{split}
{
{\cal C}_{jk}^{\rm LP}(f)=\frac{4 f_k f_j \sin ^2(\pi  f T)}{\pi ^2 T^2 \left(f^2-f_k^2\right) \left(f^2 - f_j\right)} \,,
}
\end{split}
\end{equation}
and
\begin{equation}
\label{eq:bb_filter}
\begin{split}
{
     {\cal C}_{jk}^{\rm HP}(f)=\frac{4 f^2 \sin ^2(\pi  f T)}{\pi ^2 T^2 \left( f^2 - f_k^2 \right) \left( f^2 - f_j^2 \right)} \,,
}
\end{split}
\end{equation}
which are the {transfer functions} for $\langle \alpha_{aj} \alpha_{bk} \rangle$ and $\langle \beta_{aj} \beta_{bk} \rangle$, respectively.
The relation ${\cal C}_{kj}^{\rm LP}(f)/{\cal C}_{kj}^{\rm HP}(f)\sim f^{-2}$ motivates us to refer to ${\cal C}_{kj}^{\rm LP}(f)$ and ${\cal C}_{kj}^{\rm HP}(f)$ as low-pass and high-pass {transfer functions}, respectively. These {are equivalent to the sinc-sinc function obtained in \cite{Allen:2024uqs} and} are shown in Figure \ref{fig:filters}. The above expressions can be used to similarly show that the $\alpha$- and $\beta$-bins, $\langle \alpha_{aj} \beta_{bk} \rangle$, are uncorrelated and different frequency bins ($k \neq j$) are very weakly correlated{; but not insignificant \cite{Crisostomi:2025vue}}. {The quantities $\langle \alpha_{aj} \alpha_{bk} \rangle$ and $\langle \beta_{aj} \beta_{bk} \rangle$ are the frequency- and Fourier-domain representation of the covariance function.}

{
The transfer functions (\ref{eq:aa_filter}-\ref{eq:bb_filter}) provide physical insight into the nature of the observed signal. If a stochastic signal had support at and only at the set of frequencies $\{f_k = k/T\}$, then the Fourier modes would be uncorrelated and the diagonal approximation to the covariance function would be exact. However, the SGWB is expected to have a coherence timescale $T_{\rm gw}={\cal O}\left(10^6\right)$ years. This means the SGWB is effectively continuous in the nanohertz band and the observed signal is but a short snapshot of a longer process \cite{Allen:2024uqs}. The windowing of the signal partially breaks stationarity and consequently distinguishes sines and cosines by their phase difference.
The transfer functions account for inter-frequency and off-diagonal corrections induced by the act of observation \cite{Allen:2024uqs, Crisostomi:2025vue, Bernardo:2025dat} \footnote{{Mathematically, a stationary random process has decoupled Fourier modes, equal probability distribution functions and frequency dependence between sines and cosines. An observation in the Fourier-domain with a period $T\sim T_{\rm s}$ (coherence time) will resolve these signal properties, because the signal covariance is approximately diagonalized in the basis $\Vec{B}'=\{ \sin(2\pi k t/T), \cos(2\pi k t/T) \}$.
But, the basis that exactly diagonalizes the signal covariance is $\Vec{B}=\{ \sin(2\pi k t/T_{\rm s}), \cos(2\pi k t/T_{\rm s}) \}$; $\vec{B} \neq \vec{B}'$.}}.
}

\begin{figure}[t]
    \centering    {\includegraphics[width=0.45\textwidth]{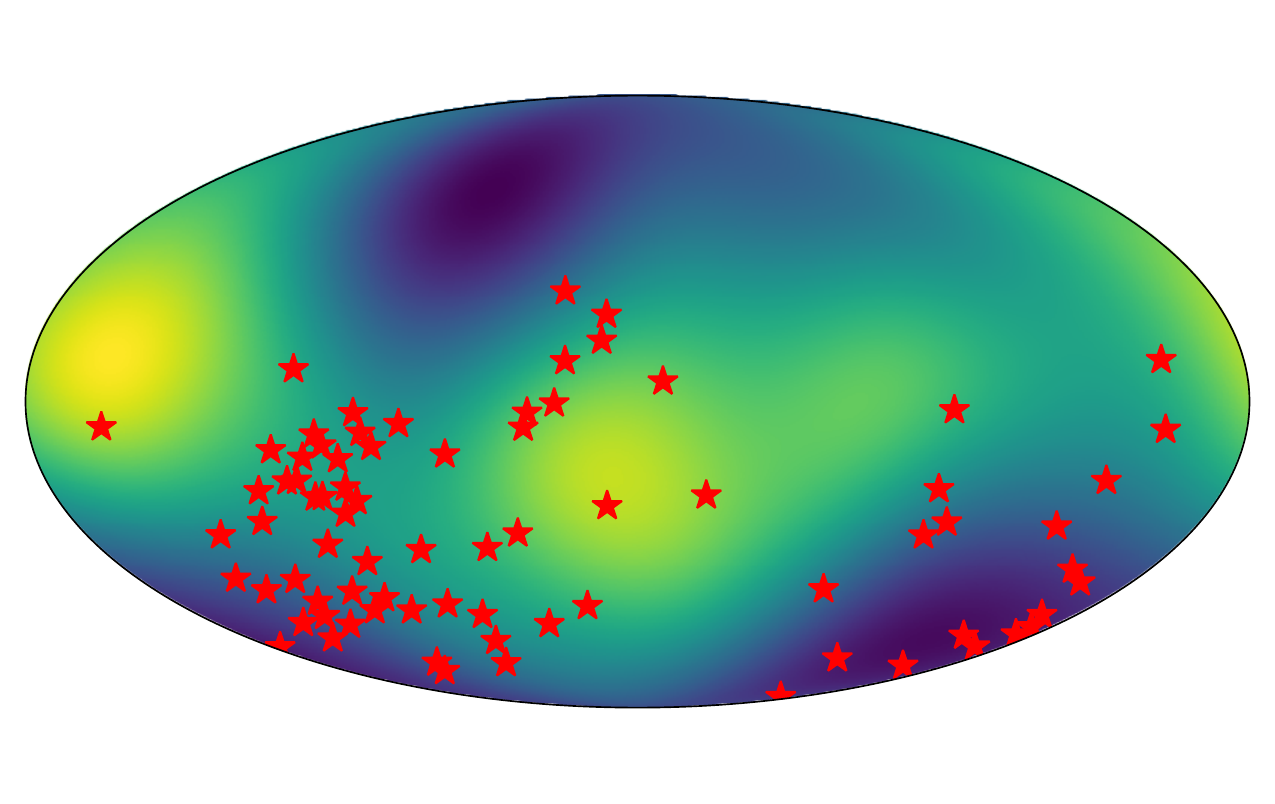}}
    \caption{Sky positions of the Meerkat PTA pulsars (red stars) \cite{Miles:2022lkg} with a smoothed projection/map of a SGWB.}
    \label{fig:mpta_gwb}
\end{figure}

To show that the above formalism {and physical interpretation} is robust, we turn to SGWB simulations in PTA using the standard \texttt{TEMPO2}~\cite{Hobbs:2006cd, Hobbs:2009yn} timing software and its python wrapper \texttt{libstempo}\footnote{https://github.com/vallis/libstempo}. We take the Meerkat PTA's 78 millisecond pulsars \cite{Miles:2022lkg} as reference positions on the sphere (Figure \ref{fig:mpta_gwb}), and simulate 15-year PTA mock data with input GWs at frequencies $f\in (1, 100)$ nHz from ${\cal O}(10^3)$ SMBHBs, a strain amplitude $A_{\rm gw}=2.4 \times 10^{-15}$, and a spectral index $\gamma_{\rm gw}=13/3$. We emphasize that our results hold irrespective of the arbitrary reference to Meerkat PTA pulsars. The sample variances of the $\alpha$- and $\beta$-bins are measured, and the ensemble statistics/cosmic variance is accounted for by repeating the simulation numerous times{; in this case, the ensemble composed of $10^3$ \texttt{TEMPO2} mock data realizations, each generated with the signal parameters $(A_{\rm gw}, \gamma_{\rm gw})$ \cite{Bernardo:2025dat}}. We then use a Markov chain Monte Carlo (MCMC) approach to compare the bin variances of the mock data, $\langle \alpha_{k}^2 \rangle_{\rm sims}$ and $\langle \beta_{k}^2 \rangle_{\rm sims}$, to models based on (\ref{eq:xy_correlation}-\ref{eq:bb_filter}) with power spectrum models $I_{{\rm M}_0}(f)=I_{\rm ref} \overline{f}^{-7/3}$ and $I_{{\rm M}_1}(f)=I_{\rm ref} \overline{f}^{2-\gamma}$ where $\overline{f}=f/(1 \ {\rm nHz})$\footnote{$I_{\rm ref}$ is the power spectrum at a frequency $\sim1$  nHz.}. {Leaving out the off-diagonal terms}, we consider a likelihood ${\cal L}$ given by $\log {\cal L} \sim -0.5 \sum_{k} \left( (W_{k, {\rm M}} - \overline{W}_{k, {\rm sims}})/\Delta W_{k, {\rm sims}} \right)^2$ where $W_{k, {\rm M}}$ and $W_{k, {\rm sims}}$ are timing residual bin variances of the model and the mock data, respectively. 

\begin{figure}[t]
    \centering
    {\includegraphics[width=0.45\textwidth]{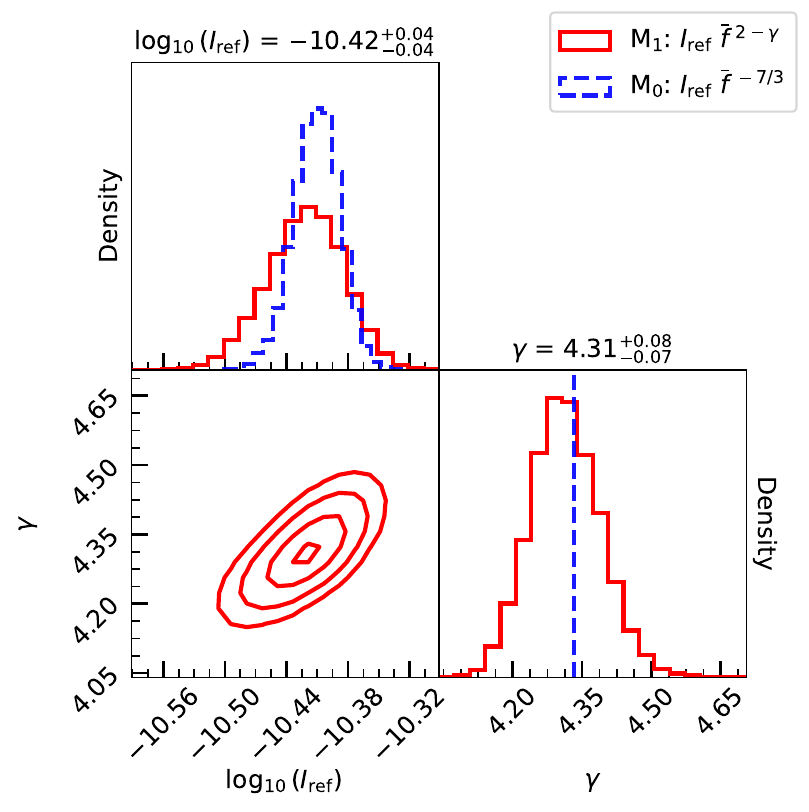}}
    \caption{Constraints on the power spectrum models $I_{{\rm M}_0}(f)=I_{\rm ref} \overline{f}^{-7/3}$ and $I_{{\rm M}_1}(f)=I_{\rm ref} \overline{f}^{2-\gamma}$ with accurate bin variances (\ref{eq:xy_correlation}-\ref{eq:bb_filter}) using 15-yr Meerkat PTA mock data with input GWs at frequencies $f\in (1, 100)$ nHz from ${\cal O}(10^3)$ SMBHBs, $A_{\rm gw}=2.4 \times 10^{-15}$, and $\gamma_{\rm gw}=13/3$.}
    \label{fig:corner_sims}
\end{figure}

\begin{figure}[t]
    \centering
    {\includegraphics[width=0.45\textwidth]{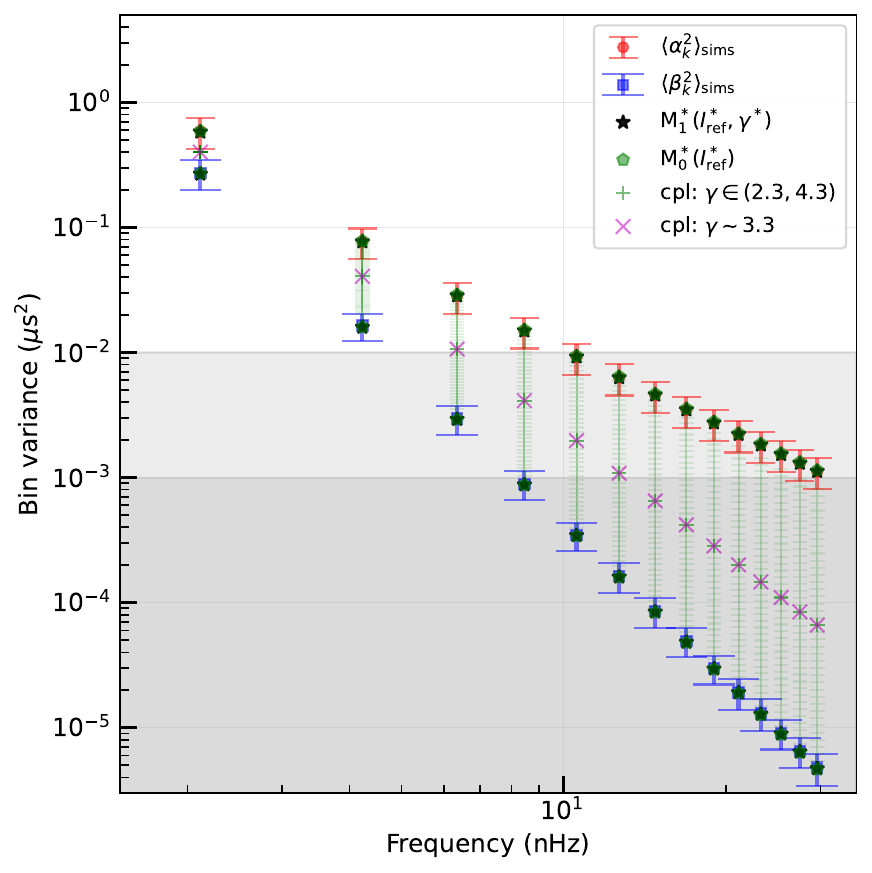}}
    \caption{Simulated bin variances in 15-yr Meerkat PTA mock data (red and blue points/error bars) with input GWs at frequencies $f\in (1, 100)$ nHz from ${\cal O}(10^3)$ SMBHBs, $A_{\rm gw}=2.4 \times 10^{-15}$, and $\gamma_{\rm gw}=13/3$. Black stars and green pentagons are best fits from models ${\rm M}_1^*(I_{\rm ref}^*, \eta^*)$ and ${\rm M}_0^*(I_{\rm ref}^*)$, respectively, and (\ref{eq:xy_correlation}-\ref{eq:bb_filter}). Violet `$\times$' and green `$+$' are bin variances corresponding to a common power law model saturated to fit both $\alpha$- and $\beta$-bin variances of the mock data. Shaded regions below thresholds $\sim10^{-3}$ $\mu$s$^2$ and $\sim10^{-2}$ $\mu$s$^2$ depict typical one-point noise levels.}
    \label{fig:bin_variance_bestfit}
\end{figure}

\begin{figure}[t]
    \centering
    {\includegraphics[width=0.45\textwidth]{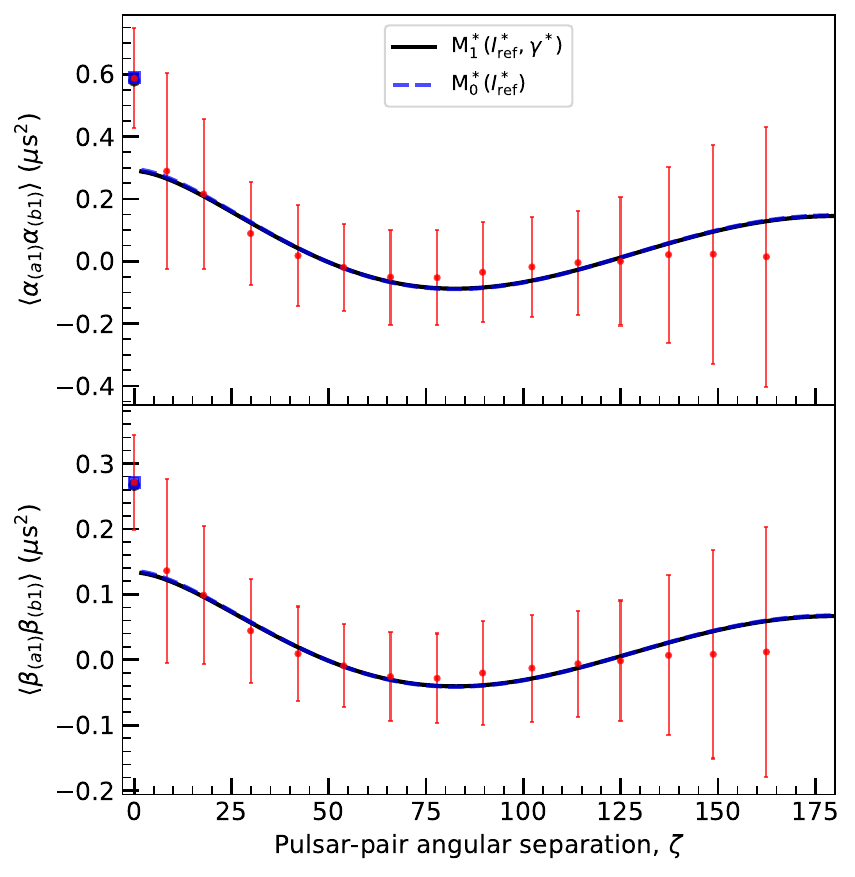}}
    \caption{Correlation in the 1st frequency bin, $f_1\sim 2$ nHz, in 15-yr Meerkat PTA mock data (red points/error bars) with input GWs at frequencies $f\in (1, 100)$ nHz from ${\cal O}(10^3)$ SMBHBs, $A_{\rm gw}=2.4 \times 10^{-15}$, and $\gamma_{\rm gw}=13/3$. Black-solid and blue-dashed lines show the best fits ${\rm M}_1^*(I_{\rm ref}^*, \eta^*)$ and ${\rm M}_0^*(I_{\rm ref}^*)$, respectively.}
    \label{fig:correlation_bestfit}
\end{figure}

The constraints are shown in Figure \ref{fig:corner_sims}, and the corresponding best fits to the bin variances and the correlation are shown in Figures \ref{fig:bin_variance_bestfit}-\ref{fig:correlation_bestfit}, demonstrating excellent performance by the model. The takeaway is that the accurate (\ref{eq:xy_correlation}-\ref{eq:bb_filter}) with a power law spectrum is naturally able to recover the precise spectral tilt, $\gamma\sim \gamma_{\rm gw}=13/3$, expected of a SGWB from a myriad of SMBHBs. Notably, both one- and two-parameter models, M$_0$ and M$_1$, are consistent with inferencing the power spectrum's reference value, $I_{\rm ref}$. This consistency between the two models can also be seen in the best fit bin variances and correlation, as shown in Figures \ref{fig:bin_variance_bestfit}-\ref{fig:correlation_bestfit}. The correlation is in tune with the HD curve, indicative of a {GW} signal. It is worth highlighting that the choice of reference frequency is immaterial, i.e., any choice gives the same posterior, with appropriately shifted amplitude\footnote{If $f_{\rm ref}\rightarrow f_{\rm ref}'$, then $I_{\rm ref}'= I_{\rm ref} \left( f_{\rm ref}'/f_{\rm ref} \right)^{2-\gamma}$. This changes the covariance between $I_{\rm ref}$ and $\gamma$; but this is due to statistics and not a physical effect.}.

However, a main point of this letter emerges illustratively through the bin variances inferred by {a} common power law (cpl), $I_{\rm cpl}(f)=I_{\rm ref} \overline{f}^{2-\gamma}$, anchored on \eqref{eq:PTA_ab_correlation} that is fitted to both $\alpha$- and $\beta$-bin variances{; in the language of this letter, cpl refers to the diagonal approximation to the covariance function}. This is shown in Figure \ref{fig:bin_variance_bestfit}, when the cpl spectral index $\gamma$ is varied in $\gamma \in (2.3, 4.3)$. A cpl based on \eqref{eq:PTA_ab_correlation} fitted to both $\alpha$- and $\beta$-bin variances always infers a bluer spectrum contrary to the expectation; for the SMBHB spectrum, cpl turns in $ 7/3 \lesssim \gamma \lesssim 13/3$. In addition to the spectral index, a cpl based on \eqref{eq:PTA_ab_correlation} is likely to output a systematic error to the inferred amplitude of the spectrum because of the power difference between the $\alpha$- and $\beta$-bins. This turns out to be less of an issue after fitting the timing model and considering one-point noise, roughly equalizing the power between bins. A systematic error and dependence on the reference frequency will persist {when} the data is sufficiently precise.

\begin{table}[t]
    \centering
    \renewcommand{\arraystretch}{1.3}
    \caption{Constraints on the power spectrum $I(f)$ using the standard analysis (\eqref{eq:PTA_ab_correlation}+cpl) and this work ((\ref{eq:xy_correlation}-{\ref{eq:bb_filter}})+M$_1$) given \texttt{TEMPO2} simulated 15-yr Meerkat PTA data with input GWs at frequencies $f\in (1, 100)$ nHz from ${\cal O}(10^3)$ SMBHBs, $A_{\rm gw}=2.4 \times 10^{-15}$, and $\gamma_{\rm gw}=13/3$. {The results are obtained using $10^3$ independent mock data realizations, each one turning in estimates for $(I_{\rm gw}, \gamma_{\rm gw})$. The numbers presented are the ensemble averages for the $10^3$ mock data realizations.}}
    \begin{tabularx}{\columnwidth}{|c|X|X|c|}
    \hline
    Model & \ Noise ($\mu$s$^2$) & \ $\log_{10}(I_{\rm ref})$ & \phantom{gggg} $\gamma$ \phantom{gggg} \\ \hline
    \multirow{3}{*}{\eqref{eq:PTA_ab_correlation}+cpl}
    & $0$ & $-10.33^{+0.06}_{-0.07}$ & $4.22^{+0.08}_{-0.09}$ \\ \cline{2-4} 
    & $10^{-3}$ & $-10.5^{+0.1}_{-2.2}$ & $3.9^{+0.4}_{-3.9}$ \\ \cline{2-4} 
    & $10^{-2}$ & $-10.6^{+0.2}_{-0.7}$ & $3.0^{+0.9}_{-2.2}$ \\ \hline
    \multirow{3}{*}{(\ref{eq:xy_correlation}-{\ref{eq:bb_filter}})+M$_1$}
    & $0$ & $-10.42\pm 0.04$ & $4.31^{+0.08}_{-0.07}$ \\ \cline{2-4} 
    & $10^{-3}$ & $-10.42^{+0.04}_{-0.05}$ & $4.2\pm0.2$ \\ \cline{2-4} 
    & $10^{-2}$ & $-10.4 \pm 0.1$ & $4.5^{+0.9}_{-0.5}$ \\ \hline
    \end{tabularx}
    \label{tab:constraints}
\end{table}

Elaborating on the pitfalls of associating the bin variances with \eqref{eq:PTA_ab_correlation} and a cpl, we consider different levels of one-point noise, emphasizing that the bin variances are one-point measures of the signal. The constraints are shown in Table \ref{tab:constraints}. In the noise-free case, depicted by the red and green points in Figure \ref{fig:bin_variance_bestfit}, a cpl turns out to be able to infer the SMBHB spectrum, but only by happenstance since the $\beta$-bins have more weight in the likelihood due to their smaller uncertainty compared to the $\alpha$-bins. It is notable that the noise-free case hints at a systematically bluer spectrum at $68\%$ confidence compared to the expected SMBHB spectrum. This systematic error is exacerbated by any level of one-point noise. At $\sim10^{-3}$ $\mu$s$^2$, disabling the sensitivity to $\beta$-bins for $f \gtrsim 10$ nHz, a cpl infers $\log_{10}(I_{\rm ref})=-10.5^{+0.1}_{-2.2}$ and $\gamma=3.9^{+0.4}_{-3.9}$. With noise at $\sim10^{-2}$ $\mu$s$^2$, the inferred parameters are $\log_{10}(I_{\rm ref})=-10.6^{+0.2}_{-0.7}$ and $\gamma=3.0^{+0.9}_{-2.2}$. Accordingly, the posterior of the amplitude and the spectral index becomes bimodal, reaching local maxima at the $\alpha$- and $\beta$-bins because \eqref{eq:PTA_ab_correlation} and cpl is saturated to fit two different {power laws}. Undesirably, a blue-tilted spectrum compared to SMBHB is always inferred by the analysis with a cpl.

In contrast, the same typical one-point noise levels do not lead to analogous systematic error contamination when using (\ref{eq:xy_correlation}-\ref{eq:bb_filter}) {(Table \ref{tab:constraints})}. The inferred parameter posteriors are consistent with each other and the expected values, only statistically wider depending on the amount of noise present.

{
We emphasize that our analysis does not use the variances as a direct estimator of the power spectrum. Instead, we compute the covariance function in the frequency- and Fourier-domain, evaluating integrals of the form \eqref{eq:F_functional}.
}
{O}ur analysis could also be extended to post-fit residuals, in which case the power spectrum in (\ref{eq:xy_correlation}-\ref{eq:bb_filter}) must be changed to $I(f) {\cal T}(f)$ where ${\cal T}(f)$ is the transmission function \cite{Hazboun:2019vhv}. {This will be pursued elsewhere.}

\begin{table}[t]
    \renewcommand{\arraystretch}{1.3}
    \caption{
    {
    Stationary correlated random processes in PTA data, their covariance function (CF), power spectral density (PSD), and bin variances. WN and RN stand for white and red noise, respectively. $S(f)$ is an arbitrary function, $\delta(x)$ is Dirac's delta function, $T$ is the observation period, $j, k, q \in \mathbb{N}$, $\{f_q\}$ is a set of signal mode frequencies, and $D_\Delta(f)$ is a peak spectrum of width $\Delta \lesssim 1/T$. The corrections correspond to inter-frequency and off-diagonal terms (see Eq.~(34) of \cite{Bernardo:2025dat}).
    The last column is a statement on the bin variances; `True' implies $\langle \alpha_k^2 \rangle = \langle \beta_k^2 \rangle$. Superscript $^\star$ denotes that the signal is correlated across pulsars.
    }
    }
    \centering
   \begin{tabularx}{\columnwidth}{|c|
   >{\hsize=0.81\hsize}X|
   >{\hsize=1.19\hsize}X
   |c|}
    \hline
    {\scriptsize Process} & CF $(\tau=t-t')$ & PSD & {\scriptsize $\alpha=\beta$} \\ \hline
    
    WN & $\delta(\tau)/2$ & $1$ & True \\ \hline
    
    \multirow{3}{*}{RN} 
     & { \scriptsize $\sum_{k} S(f_k)\cos\left(\tfrac{2\pi k \tau}{T}\right)$ }
     & { \scriptsize $\sum_k S(f)\,\delta\left(f-\tfrac{k}{T}\right)$ }
     & True \\ \cline{2-4}
    
     & { \tiny $\sim\sum_{j} S(f_{j})\cos\left(\tfrac{2\pi j \tau}{T}\right)$ }
     & { \tiny $\sum_q S(f)\,D_\Delta\left(f-f_q\right)\big|_{\Delta\lesssim\frac{1}{T}, f_q\sim\frac{j}{T}}$ }
     & True \\ \cline{2-4}
    
     & { \tiny $\sim\sum_q S(f_q)\cos\left(2\pi f_q \tau\right)$ }
     & { \tiny $\sum_q S(f)\,D_\Delta\left(f-f_q\right)\big|_{\Delta > \frac{1}{T}}$ }
     & False \\ \hline
    
    {\scriptsize SGWB$^{\star}$} & {\scriptsize $\int df\, S(f)\cos(2\pi f \tau)$ } & $S(f)$ & False \\ \hline
    
    \end{tabularx}
    \label{tab:correlated_processes}
\end{table}

{Lastly, while our letter focused on the SGWB signal, the formalism accommodates general stationary correlated random process, such as white and red noises in PTA analysis. Table \ref{tab:correlated_processes} gives the relation between the time-domain covariance function, the power spectral density, and the Fourier bin variances. White noise is a special case, uncorrelated in time and frequency. Pulsar red noise can be understood either with infinitesimally narrow or finite width power spectrum at the observational frequencies, corresponding to whether the Fourier coefficients will have equal or unequal bin variances. The transfer functions (\ref{eq:aa_filter}-\ref{eq:bb_filter}) give an extra layer of information for characterizing the SGWB signal and colored noises in PTA data.}

We highlight the roles of the temporal correlation \eqref{eq:tt'_correlation}, and the corresponding {transfer functions} (\ref{eq:aa_filter}-\ref{eq:bb_filter}), in our analysis, which we deem are as profound as the HD curve for recognizing a SGWB signal. In this letter, we have shown that PTA residual variances and correlation based on (\ref{eq:xy_correlation}-\ref{eq:bb_filter}) align perfectly with simulated signal from SMBHBs. Naturally, the accuracy of the expressions used for interpretation must meet the precision of the data in order to extract meaningful insights. Our results have shown that \eqref{eq:PTA_ab_correlation} {or a diagonal approximation of the covariance function} used to interpret a SGWB signal inevitably introduce{s} systematic errors in the inferred spectrum (Table \ref{tab:constraints}). More sophisticated models may be able to get away with a better fit, but this would not help out in the physical interpretation of the signal. Ultimately, the SGWB piece of the covariance of the PTA likelihood must be updated according to (\ref{eq:xy_correlation}-\ref{eq:bb_filter}) to properly account for these systematic biases when analyzing the data {in the frequency- and Fourier-domain} \cite{Valtolina:2024kdb, Crisostomi:2025vue}. This shows that an inferred bluer signal can be a byproduct of such systematics \cite{Allen:2024uqs,Crisostomi:2025vue} and potentially hints at a conservative solution to the debated source of the PTA signal, tracing it to astrophysics.

The authors thank Achamveedu Gopakumar, Subhajit Dandapat, and Debabrata Deb for relevant discussion on PTA simulations{, as well as Bruce Allen, Rutger Van Haasteren, and Boris Goncharov for many important discussions on PTA analysis}. {RCB would also like to acknowledge the kind hospitality of Institute of Physics, Academia Sinica in Taiwan.}
RCB was supported by an appointment to the JRG Program at the APCTP through the Science and Technology Promotion Fund and Lottery Fund of the Korean Government, and was also supported by the Korean Local Governments in Gyeongsangbuk-do Province and Pohang City. This work was supported in part by the National Science and Technology Council of Taiwan, Republic of China, under Grant No. NSTC 113-2112-M-001-033. 



%

\end{document}